%% Use Tex. Macro harvmac.

\input harvmac
\input amssym.def
\input epsf
\noblackbox
%%% Paragraphs
\newcount\figno
\figno=0
\def\fig#1#2#3{
\par\begingroup\parindent=0pt\leftskip=1cm\rightskip=1cm\parindent=0pt
\baselineskip=11pt
\global\advance\figno by 1
\midinsert
\epsfxsize=#3
\centerline{\epsfbox{#2}}
\vskip 12pt
\centerline{{\bf Figure \the\figno} #1}\par
\endinsert\endgroup\par}
\def\figlabel#1{\xdef#1{\the\figno}}
\def\pano{\par\noindent}

%%% special math symbols
\def\pmb#1{\setbox0=\hbox{#1}%
 \kern-.025em\copy0\kern-\wd0
 \kern.05em\copy0\kern-\wd0
 \kern-.025em\raise.0433em\box0 }
\font\cmss=cmss10
\font\cmsss=cmss10 at 7pt
\def\inbar{\,\vrule height1.5ex width.4pt depth0pt}
\def\half{{1\over 2}}
\def\rlx{\relax\leavevmode}
\def\Cop{\relax\,\hbox{$\inbar\kern-.3em{\rm C}$}}
\def\Rop{\relax{\rm I\kern-.18em R}}
\def\Nop{\relax{\rm I\kern-.18em N}}
\def\one{\relax{\rm 1\kern-.25em I}}
\def\Pop{\relax{\rm I\kern-.18em P}}
\def\Zop{\rlx\leavevmode\ifmmode\mathchoice{\hbox{\cmss Z\kern-.4em Z}}
 {\hbox{\cmss Z\kern-.4em Z}}{\lower.9pt\hbox{\cmsss Z\kern-.36em Z}}
 {\lower1.2pt\hbox{\cmsss Z\kern-.36em Z}}\else{\cmss Z\kern-.4em
 Z}\fi}

%%% misc.

\def\N{{\cal N}}

\def\H{{\cal H}}
\def\I{{\cal I}}

\def\half{{1\over 2}}
\def\ie{{\it i.e.}}
\def\eg{{\it e.g.}}

\def\sl2r{SL_2\Rop}

\def\g{{\frak g}}

\def\h{{\frak h}}

\def\arrow{\rightarrow}

%%%%

%\draftmode

%%% further macros

\lref\FHTtwo{D.S. Freed, {\it The Verlinde algebra is twisted
equivariant K-theory}, Turkish J. Math. {\bf 25} no. 1, 159 (2001);
{\tt math.RT/0101038}.}

\lref\FHTone{D.S. Freed, M.J. Hopkins, C. Teleman, {\it Twisted
equivariant K-theory with complex coefficients}, {\tt math.AT/0206257}.}      

\lref\FHTthree{D.S. Freed, M.J. Hopkins, C. Teleman, {\it Twisted
K-theory and the Verlinde algebra}, in preparation.}

\lref\FHTintegral{D.S. Freed, M.J. Hopkins, C. Teleman, {\it Twisted K
theory and Loop group representations}, to appear. }

\lref\MooreMinasian{
R.~Minasian and G.~W.~Moore, {\it K-theory and Ramond-Ramond charge},
JHEP {\bf 9711}, 002 (1997); {\tt hep-th/9710230}.}

\lref\WittenK{
E.~Witten, {\it D-branes and K-theory},
JHEP {\bf 9812}, 019 (1998); {\tt hep-th/9810188}.
}

\lref\mrgnonbps{
M.~R.~Gaberdiel, {\it Lectures on non-BPS Dirichlet branes},
Class.\ Quant.\ Grav.\  {\bf 17}, 3483 (2000); {\tt hep-th/0005029}.}

\lref\FSWZW{
S.~Fredenhagen, V.~Schomerus,
{\it Branes on group manifolds, gluon condensates, and twisted K-theory},
JHEP {\bf 0104}, 007 (2001);
{\tt hep-th/0012164}.
}

\lref\FSdynamics{
S.~Fredenhagen and V.~Schomerus,
{\it Brane dynamics in CFT backgrounds};
{\tt hep-th/0104043}.}

\lref\MMSII{
J.~Maldacena, G.~Moore, N.~Seiberg, {\it D-brane instantons and
K-theory charges}; {\tt hep-th/0108100}.
}

\lref\volkerK{V. Braun, {\it  Twisted K-Theory of Lie Groups}; {\tt hep-th/0305178}. }

\lref\GKO{P. Goddard, A. Kent, D. Olive, {\it Unitary Representations
 Of The Virasoro And Supervirasoro Algebras}, Commun.\ Math.\ Phys.\
 {\bf 103}, 105 (1986).}

\lref\KS{Y.~Kazama, H.~Suzuki,
{\it New N=2 Superconformal Field Theories And Superstring Compactification},
Nucl.\ Phys.\ B {\bf 321}, 232 (1989).}

\lref\FZ{
A.~B.~Zamolodchikov, V.~A.~Fateev,
{\it Operator Algebra And Correlation Functions In The Two-Dimensional Wess-Zumino SU(2) X SU(2) Chiral Model},
Sov.\ J.\ Nucl.\ Phys.\  {\bf 43}, 657 (1986)}

\lref\gepnerFI{
D.~Gepner, {\it Field Identification In Coset Conformal Field Theories},
Phys.\ Lett.\ B {\bf 222}, 207 (1989).}

\lref\horivafa{
K.~Hori, A.~Iqbal, C.~Vafa, {\it D-branes and mirror symmetry}; {\tt hep-th/0005247}.
}

\lref\kac{V. Kac,
{\it Infinite Dimensional Lie Algebras}, Cambridge University Press, (1990).}

\lref\walton{
M.~A.~Walton,
{\it Fusion Rules In Wess-Zumino-Witten Models},
Nucl.\ Phys.\ B {\bf 340}, 777 (1990).}

\lref\MMSI{
J.~Maldacena, G.~Moore, N.~Seiberg, {\it Geometrical interpretation
of D-branes in gauged WZW models}, JHEP {\bf 0107}, 046 (2001); {\tt hep-th/0105038}.}

\lref\mooreclay{G.~Moore, {\it Lectures on Branes, K-Theory and RR
Charges}, Clay Mathematical Institute Lectures;
{\tt http://www.physics.rutgers.edu/$\tilde{\ }$gmoore/clay.html}.}

\lref\LW{W.~Lerche, J.~Walcher, {\it Boundary rings and N = 2 coset models},
Nucl.\ Phys.\ B {\bf 625}, 97 (2002); {\tt hep-th/0011107}.}

\lref\stefanBF{
S.~Fredenhagen,
{\it Organizing boundary RG flows}; {\tt hep-th/0301229}.}

\lref\douglasfiol{
M.~R.~Douglas, B.~Fiol, {\it D-branes and discrete torsion. II}, {\tt hep-th/9903031}.}

\lref\BDLR{I.~Brunner, M.~R.~Douglas, A.~E.~Lawrence, C.~R\"omelsberger,
{\it D-branes on the quintic}, JHEP {\bf 0008}, 015 (2000); {\tt hep-th/9906200}.}

\lref\LVW{W.~Lerche, C.~Vafa, N.~P.~Warner, {\it Chiral Rings In N=2
 Superconformal Theories}, Nucl.\ Phys.\ B {\bf 324}, 427 (1989).}

\lref\FS{S.~Fredenhagen, V.~Schomerus,
{\it D-branes in coset models}, JHEP {\bf 0202}, 005 (2002); {\tt hep-th/0111189}. }

\lref\SegalKG{G. Segal, {\it Equivariant K-theory}, Inst. Hautes
\'Etudes Sci. Publ. Math. {\bf 34}, 129 (1968).}

\lref\atiyahsegal{
M.F. Atiyah, G.B. Segal, 
{\it Equivariant K-theory, and completion}, 
J. Differential Geometry {\bf 3}, 1 (1969).}

\lref\CBMS{J.~P.~May, {\it Equivariant Homotopy and Cohomology
Theory}, CBMS, Regional Conference Series {\bf 91}, AMS, 1996.}

\lref\atiyah{
M.F. Atiyah, {\it K-theory Past and Present}, {\tt math.KT/0012213}}

\lref\freedloops{
D.~S.~Freed, {\it Twisted K-theory and loop groups};
{\tt math.at/0206237}.}

\lref\Steenrod{N.E. Steenrod, {\it Cohomology operations}, Annals of
Mathematics Studies {\bf 50}, Princeton University Press (1962).}

\lref\KT{V.~G.~Kac, I.~T.~Todorov,
{\it Superconformal Current Algebras And Their Unitary Representations},
Commun.\ Math.\ Phys.\  {\bf 102}, 337 (1985). }

\lref\schnitzerlevelrank{
S.~G.~Naculich, H.~J.~Schnitzer,
{\it Superconformal coset equivalence from level-rank duality},
Nucl.\ Phys.\ B {\bf 505}, 727 (1997); {\tt hep-th/9705149}.}

\lref\hodgkin{L. Hodgkin, {\it The equivariant K\"unneth theorem in
K-theorem}, in {\it Topics in K-theory. Two independent
contributions}, Lecture Notes in Math. {\bf 496}, Springer, Berlin, 1975. }

\lref\snaith{V.P. Snaith, {\it On the K\"unneth formula spectral
sequence in equivariant $K$-theory}, Proc. Cambridge Philos. Soc. {\bf
72}, 167, (1972).}

\lref\pittie{H.V. Pittie, {\it Homogeneous vector bundles on
homogeneous spaces}, Topology {\bf 11}, 199 (1972).}

\lref\gepnerFRG{D.~Gepner, {\it Fusion Rings and Geometry}, {\it Fusion Rings And Geometry},
Commun.\ Math.\ Phys.\  {\bf 141}, 381 (1991).
}

\lref\FH{W. Fulton, J. Harris, {\it Representation Theory}, Springer, 1991.}

\lref\PS{A. Pressley, G. Segal, {\it Loop Groups}, Oxford, 1986.}

\lref\Hartshorn{R.~Hartshorne, {\it Algebraic Geometry}, Springer, 1977.}

\lref\adams{J.F. Adams, {\it Vector Fields on Spheres}, Ann. of
 Math. {\bf 74}, 603 (1962).}

%%%%%%%%%%%%%%%%%%

\lref\WittenGrassmann{E. Witten, {\it The Verlinde Algebra And The
Cohomology Of The Grassmannian}, {\tt hep-th/9312104}.}

%%% Title page

\Title{\vbox{
\hbox{hep-th/0308058}
\hbox{DAMTP-2003-76}}}
{\vbox{\centerline{D-branes in ${\cal N}=2$ coset models and}
\smallskip
\smallskip
\centerline{twisted equivariant K-theory}}}
\centerline{
Sakura Sch\"afer-Nameki\ \footnote{$^{\sharp}$}{{\tt e-mail: 
S.Schafer-Nameki@damtp.cam.ac.uk}}}
\bigskip
\centerline{\it Department of Applied Mathematics and Theoretical
Physics}  
\centerline{\it University of Cambridge, Wilberforce Road, Cambridge CB3 OWA, U.K.}
\smallskip
\vskip2cm
\centerline{\bf Abstract}
\bigskip
\noindent
The charges of D-branes in Kazama-Suzuki coset models are analyzed. 
We provide the calculation of the corresponding twisted equivariant
K-theory, and in the case of Grassmannian cosets, $su(n+1)/u(n)$,
compare this to the charge lattices that are derived
from boundary conformal field theory. 

\bigskip

\Date{8.8.2003}

%%%%

\newsec{Introduction}

\noindent
Twisted K-theory has recently attracted much attention, both in
mathematics and in string theory. Most notable among the results
obtained in this subject is 
the theorem by Freed, Hopkins and Teleman (FHT) relating twisted K-theory to the Verlinde
algebra \refs{\FHTtwo, \FHTthree,\FHTone}, which seen in the light of
efforts in string theory to relate K-theory to D-brane charges, as
computed from conformal field theory data, is indeed very suggestive.

In string theory, the importance of K-theory is
founded upon the conjecture that D-brane charges are classified by the K-theory of the target space
\refs{\MooreMinasian, \WittenK}. This conjecture has been tested in
various instances, when both an algebraic description of the D-branes in terms of
boundary conformal field theory (BCFT) as well as the relevant K-theory are accessible to explicit
computations. Most prominent examples of theories, where the BCFT is
under control are free CFTs, orbifolds, WZW and coset
models. The relation to K-theory has been established in a vast number
of examples for the free and orbifold theories. In the case of
(supersymmetric) WZW models, the presence of the non-trivial
background B-field implies that D-brane charges should take values
in a twisted version of K-theory. 
The K-theories relevant for WZW models have been computed in
\refs{\FSWZW, \MMSII, \volkerK} and indicate that the known boundary
states do not in general suffice to fill up the charge lattice that is
predicted by K-theory.

In this paper the correspondence between
D-brane charges and K-theory shall be explored for coset CFTs
\refs{\GKO}. Supersymmetry is key to the present
discussion, and the relevant models are the Kazama-Suzuki coset theories
\refs{\KS}, of which the Grassmannian and generalized parafermionic
cosets shall be of foremost interest. Using the FHT result, we
provide the computation of the twisted equivariant K-groups associated
to these models and discuss the relation to the BCFT. 

The plan of the paper is as follows. In section 2, we discuss as a
prelude the super-minimal models, and compare K-theory to BCFT in this
simplest example. 
Section 3 provides some detail on the CFT-side of the cosets. 
The main result is contained in section 4, where we use the theorem by FHT
\refs{\FHTtwo, \FHTthree} to compute the twisted equivariant
K-theory of the coset models. A matching with the BCFT results is established in the case
of Grassmannian cosets, and we close with concluding remarks and acknowledgements.

%%%%

\newsec{Superparafermions}

\noindent
The super-minimal models, realized as superparafermionic $u(1)$ cosets of the $su(2)_k$
WZW models, are the simplest examples of ${\cal N}=2$ superconformal
field theories. In this section we review the boundary CFT in
these models, as well as compute the
K-theory lattice of D-brane charges. This should serve as an appetizer
to subsequent sections, where generalizations to other ${\cal N}=2$
coset theories shall be discussed. 

\subsec{The boundary CFT side}

\noindent
The superparafermions
\refs{\FZ, \KS} have a representation in terms of the coset models
\eqn\SPF{
{{su}(2)_{k}\, \oplus \, u(1)_4
\over {u}(1)_{2(k+2)} } \,.}
The highest weight representations will be labeled by
$(\Lambda, m, a)$, where $\Lambda$ is an $su(2)_k$ weight, $m$ labels
the $u(1)_{2(k+2)}$ representations and $a$ specifies the weight of
$u(1)_4$. The induced level $2 \kappa=2( k+2)$ of the $u(1)_{2(k+2)}$ denominator theory
indicates that the weight $m$ takes values in $\Zop/2(k+2)$. The
theories \SPF\ are obtained as supersymmetric generalizations of the coset
construction for $su(2)/u(1)$, where $dim(su(2))- dim(u(1))$ adjoint fermions, described by
the $u(1)_4$ factor, are introduced. The corresponding representation
label $a$ takes values in $\Zop/4$. The coset selection and identification rules are
\eqn\cosetrules{\eqalign{
& \Lambda + m+a\,  \in 2\Zop\,,\cr
& (\Lambda, m, a) \equiv (k-\Lambda, m+k+2, a+2)\,,
}}
which are obtained in the standard fashion by acting with the proper
automorphism $\sigma: \Lambda =[\lambda_1] \arrow
[k- \lambda_1]$ on the modular S-matrices \refs{\gepnerFI}.
The theories \SPF\ are the simplest examples of the generalized
parafermion cosets $su(n+1)_k/u(1)^n$, and the supersymmetric Grassmannian
coset models $su(n+1)_k/u(n)$.  

The boundary states for \SPF\ are well known \refs{\horivafa, \MMSI, 
\LW, \stefanBF} and a brief review of the results necessary
for the present discussion is in place. We shall focus on A-type
boundary states for the ${\cal N}=2$ algebra. 
In view of the comparison to K-theory, the
lattice of RR-charges will be of main interest. In the BCFT this
is computed by the intersection index \refs{\douglasfiol}, which in
the case of the ${\cal N}=2$ superparafermion coset models has been obtained in
\refs{\horivafa, \BDLR, \LW}. 
Recall that the intersection index in the closed string picture is defined in
terms of the overlaps of the RR-sectors of the boundary states \refs{\BDLR, \LW}. The key
observation in \refs{\LW} is that the states with $\Lambda=0$ form a
generating set for the charge lattice of boundary states in the
superparafermion models. In detail,
the boundary state intersection for Cardy states $|\!| (\Lambda_i, m_i, 0)\rangle\!\rangle$ is
\eqn\BFR{
\left( {\cal I}_{(\Lambda_1, \Lambda_2)} \right)_{m_1, m_2} 
= N_{\Lambda_1, \Lambda_2}^{m_1-m_2} 
=\left(
\sum_{\Gamma} N_{\Gamma,
\Lambda_2}^{\Lambda_1}  \,\sum_m\,  \sum_{w\in W_{su(2)}} \epsilon_{w}
g^{-m+m_0}\right)_{m_1, m_2}\,,
}
where the first sum extends over all $su(2)$ level $k$
representations, and the second sum is over all states, that are coset
fields representing Ramond sector ground states, \ie, which by \refs{\LVW}
take the form $(\Lambda, m, a)$, with $m, \Lambda\in \Zop$, such that 
\eqn\Rgroundstates{
m= w(\Lambda+1)\,,
}
for some element $w$ in the $su(2)$ Weyl group, $W_{su(2)}= \langle\pm
\one\rangle$. Further
$m_0=1$ is the $u(1)$ charge of the spectral flow operator, and $N$ and
$g$ are the $su(2)$ and $u(1)$ fusion matrices,
respectively. In detail, the fusion matrix element for the $u(1)$
fields $m, m_1, m_2$ is given by $(g^m)_{m_1 m_2}= \delta_{m_1, m_2+m}$, 
or equivalently, $g$ is the shift matrix $g_{m_1  m_2}=
\delta_{m_1 ,m_2+1}$.
In \BFR\ we have fixed $a=0$ and view $\I$ as a matrix in the denominator
$u(1)$ weights, $m_i$.
By performing the sum over the $u(1)$
weights $m$, that are allowed by \Rgroundstates\ for a given $su(2)$
weight $\Gamma$ we arrive at
\eqn\BFRexplicit{
{\cal I}_{(\Lambda_1, \Lambda_2)} = \sum_{\Gamma} N^{\Lambda_1}_{\Gamma,
\Lambda_2} (g^{-\Gamma} - g^{\Gamma+2})\,.
}
To prove that the $\Lambda =0$ states, with
fixed value of $a$, form a basis of the charge lattice, first note that
\eqn\zeroL{
{\cal I}_{(0,0)}= \one - g^2\,.
}
The matrices \BFRexplicit\ with general values for $\Lambda_1$ and
$\Lambda_2$ can be obtained from the following linear combination
\eqn\zerotononzeroL{\eqalign{
&(g^{-\Lambda_1} + g^{-\Lambda_1+2}+ \cdots +g^{\Lambda_1}  )\, {\cal I}_{(0,0)}\, (g^{-\Lambda_2}+g^{-\Lambda_2 +2}+
\cdots + g^{\Lambda_2}  )  \cr
&\qquad \qquad\qquad \qquad\qquad =  (g^{-\Lambda_1} + g^{-\Lambda_1+2}+ \cdots +g^{\Lambda_1}  )
\sum_\Gamma N^0_{\Gamma, \Lambda_2}\, (g^{-\Gamma } - g^{\Gamma +2})\cr
&\qquad \qquad\qquad \qquad\qquad = \sum_\Gamma N^{\Lambda_1}_{\Gamma, \Lambda_2} \, (g^{-\Gamma}- g^{\Gamma+2})\,,
}}
where the last step follows by inserting the explicit form of the
fusion coefficients (\eg \refs{\kac, \walton}). In particular for
$su(2)$ these are
\eqn\kacwalton{
N^{\Lambda_1}_{\Gamma, \Lambda_2} = \sum_{\Lambda \in P_{\Lambda_1}}
\sum_{w\in W_{su(2)}} \epsilon_w \delta_{w( \Lambda + \Gamma+1) -
\Lambda_2 -1} \,,
}
where $P_{\Lambda}$ is the weight system of $\Lambda$.
Thus all boundary intersections can be written in terms of the
intersection of $\Lambda=0$ states. \zeroL\ and the
explicit form of the matrix $g$, which shifts every $u(1)$ label by
one, implies that the rank of ${\cal I}_{(0,0)}$ is $k+1$.

There are two types of A-branes, which satisfy the gluing conditions
$G^{\pm}=\pm \bar{G}^{\mp}$, which were refered to as even (odd)
branes in \refs{\MMSI}. Naively, each of these give rise to a charge lattice $\Zop^{k+1}$, thus
resulting in the total charge lattice $\Zop^{2(k+1)}$, which is isomorphic to the Verlinde
fusion ring of $u(1)_{2(k+1)}$, \ie
\eqn\fusionringischargelattice{
V_{u(1)_{2(k+1)}} = {R_{u(1)}\over {I}_{2(k+1)}} = \Zop^{2(k+1)}\,,
} 
where $R_{u(1)}\cong \Zop [\zeta,\zeta^{-1}]$ is the representation ring of
$u(1)$ and ${I}_{2(k+1)}$ is the Verlinde ideal.

Taking the point of view of the gauged WZW model, the $\Lambda=0$ states can be
seen to form a basis for the charge lattice as follows. As explained
in \refs{\FSdynamics, \MMSI},
the parafermions \SPF\ have a disk target space (together with a
non-trivial B-field and dilaton). For fixed level $k$ we
introduce $k+2$ even and $k+2$ odd points on the boundary of the
disk. Then it was shown in \refs{\MMSI} that there are $(k+2)(k+1)$ even (odd) A-branes, corresponding
to D1-branes stretching between the even (odd) points.
The $\Lambda=0$ D1-branes $|\!| 0, n, s\rangle\!\rangle$ stretch between consecutive points (of same
chirality, that is, even or odd), \ie, between $(n-1)\pi/(k+2)$ and $(n+1)\pi/(k+2)$, and
are thus the shortest D1-branes. From the tachyons analysis in \refs{\MMSI}
we follow that two consecutive such D1-branes $|\!| 0, n,
s\rangle\!\rangle$ and $|\!| 0, n+2, s\rangle\!\rangle$, decay and merge to give
a D1-brane with non-trivial $\Lambda$, $|\!| 1, n+1,
s\rangle\!\rangle$, as depicted in Figure 1. Iterating this process we
can produce all branes (for fixed chirality) from the $\Lambda=0$
states, as expected from the above intersection index
computation. These decays have been analyzed also in \refs{\FS, \stefanBF}.

\fig{Decay of two $\Lambda=0$ into a $\Lambda=1$ D-brane for
$k=6$.}{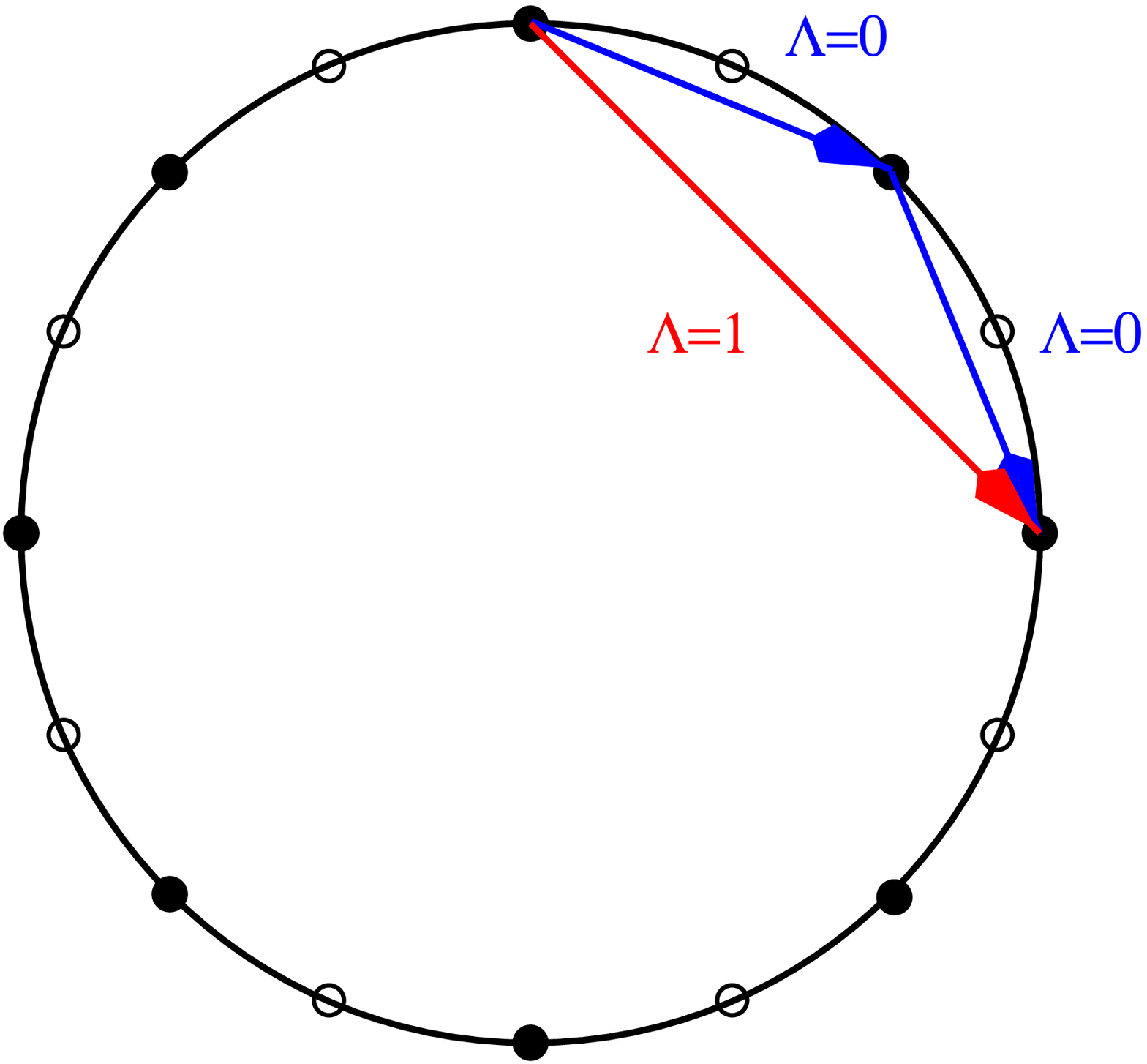}{2.2truein} 
A simple counting argument then
provides the charge lattice: there are $k+2$ even points,
and therefore $k+2$ D1-branes with $\Lambda=0$. Since in addition the
full ring of $\Lambda=0$ states is trivial, one of these branes is
redundant. The charge lattice for the even branes is therefore
$\Zop^{k+1}$. The same argument holds for the odd branes, so that the
total charge lattice is $\Zop^{2(k+1)}$. 
This discussion is therefore in agreement with the
calculation in terms of the intersection index.

\subsec{Twisted equivariant K-theory ${}^{\tau}\! K_{U(1)}(SU(2))$}

This section provides a comparison of the D-brane
charges, as computed in the last section, with K-theory. The gauged WZW
description dictates what the relevant K-theory for these backgrounds
is. The gauging implies that the K-theory needs to be equivariant with
respect to the conjugation action, and the presence of the non-trivial B-field
necessitates a non-trivial twisting with respect to a 3-form.
For other discussion of equivariant and twisted K-theory see \eg\
\refs{\SegalKG, \atiyahsegal, \CBMS, \atiyah, \mooreclay, \FHTone, \FHTtwo,\freedloops}.

In the case of the $SU(2)/U(1)$ coset we can compute the twisted
equivariant K-group directly, using a Mayer-Vietoris argument. This
calculation generalizes in principle to $SU(N)$ with
$N>2$, but the story gets more involved due to the rather complicated cell
structure (see \eg\ \refs{\Steenrod}) and other ways of computation will be more suitable. 
A calculation similar in spirit to the one presented in this section can be found in \refs{\FHTtwo} for ${}^{\tau}\!K_{SU(2)}(SU(2))$.

The geometrical action corresponding to the CFT coset is the
conjugation action. Thus, the action of $U(1)$ on $SU(2)$ is
\eqn\uoneaction{
SU(2)\ni g \,\mapsto\,  \hbox{diag}(e^{i t}, e^{-i t})\ g \ \hbox{diag}(e^{-i t}, e^{i t})\,.
}
The fixed point set is an $S^1$.
We choose an open covering of $SU(2)\sim S^3$ that is invariant under this action
\eqn\opencovers{
A\,=\, SU(2)\backslash \{-id\} \,,\qquad B\,=\, SU(2)\backslash\{id\}\,,
}
where $A$ and $B$ intersect in the equatorial $S^2$. The Mayer-Vietoris sequence for this CW triad then reads
\eqn\MVone{
\matrix{
& {}^{\tau}\! K_{U(1)}^0(SU(2)) & \longrightarrow & {}^{\tau}\!
K_{U(1)}^0(A) \oplus {}^{\tau}\! K_{U(1)}^0(B) &\longrightarrow^{^{\hskip-10pt \varphi \hskip8pt}}& {}^{\tau}\! K_{U(1)}^0(A\cap B) \cr    
& \uparrow            &&&& \downarrow \cr       
&{}^{\tau}\! K_{U(1)}^1(A\cap B) &\longleftarrow & {}^{\tau}\!
K_{U(1)}^1(A) \oplus {}^{\tau}\! K_{U(1)}^1(B)  &\longleftarrow& {}^{\tau}\! K_{U(1)}^1(SU(2))
}\,.}
Since the fixed $S^1$ intersects $A$ as well as $B$ in a point, the
corresponding groups are simply given by the representation ring of
$U(1)$, \ie
\eqn\Kone{
{}^{\tau}\! K_{U(1)}^0(A)\,=\,{}^{\tau}\! K_{U(1)}^0(B)\,=\, R_{U(1)}\,, \qquad 
{}^{\tau}\! K_{U(1)}^1(A)\,=\,{}^{\tau}\! K_{U(1)}^1(B) \,=\, 0\,.
}
To compute ${}^{\tau}\! K_{U(1)}^*(A\cap B)$ we need to apply a second
Mayer-Vietoris argument. So, choosing the analogous covering for
$A\cap B=S^2$ as in \opencovers, we obtain
\eqn\MVtwo{
\matrix{
& {}^{\tau}\! K_{U(1)}^0(S^2) & \longrightarrow & R_{U(1)}\oplus R_{U(1)}
&\longrightarrow^{^{\hskip-10pt \psi \hskip8pt}}& {}^{\tau}\! K_{U(1)}^0(S^1) \cr 
& \uparrow            &&&& \downarrow \cr       
& 0 &\longleftarrow & 0  &\longleftarrow&{}^{\tau}\!  K_{U(1)}^1(S^2)
}\,.}
Now note that $S^1= U(1)/\Zop_2$. To see this, note that the $S^1$ in
the intersection of the covers of $S^2$ is embedded into the $SU(2)$ by
\eqn\soneembedding{
S^1 \cong \left\langle \pmatrix{0 & e^{i\theta} \cr -e^{-i\theta} & 0}\right\rangle\,, 
}
whereupon the action of the $U(1)$ as of \uoneaction\ is by twice the
angle. Thus
\eqn\keytwo{
{}^{\tau}\! K_{U(1)}^0(S^1)= {}^{\tau}\! K_{\Zop_2}^0(pt.)=  \Zop^2 \,,
} 
the map $\psi$ is given by
\eqn\psimap{
\psi:\quad (\chi, \chi') \mapsto  (\chi(1)- \chi'(1)\,,\  \chi(-1)-\chi'(-1)) 
\,.}
Hence the cokernel is ${}^{\tau}\! K_{U(1)}^1(A\cap B)=0$ and the kernel is
\eqn\Kstwo{
{}^{\tau}\! K_{U(1)}^0(A\cap B) \,=\, (\,R_{U(1)}\oplus
R_{U(1)}\,)/\psi \,,
}
\ie, these are pairs of characters identified at $id$ and $-id$,
and so the quotient by $\psi$ removes two copies of $\Zop$. 

Putting this together, the map $\varphi$ in \MVone\ becomes
\eqn\psimap{\eqalign{
\varphi:\quad R_{U(1)}\oplus R_{U(1)} & \arrow \ (\,R_{U(1)}\oplus R_{U(1)}\,)\cr
(f\,,\ g)& \mapsto\ (f+gL^\kappa\,,\ f+gL^{-\kappa}) \,,
}}
where $L$ is the generator of $R_{U(1)}$ and we have used the fact
that due to the twisting, the bundles over $A$ and $B$ are patched
together by tensoring with the twist-class. So this has trivial
kernel, \ie,  ${}^{\tau}\!  K_{U(1)}^0(SU(2))=0$. The cokernel,
(without the $\psi$ identification) would be $(R_{U(1)}\oplus
R_{U(1)})/I_\kappa$, where the ideal is $I_\kappa=(L^\kappa-L^{-\kappa})$,
and so the cokernel would be isomorphic to $\Zop^{2\kappa}$. Including the $\psi$ identification
then results in $\Zop^{2\kappa-2}$. We conclude thereby that
\eqn\Ketsutwo{
{}^{\tau}\! K_{U(1)}^1(SU(2))\,=\, \Zop^{2\kappa-2}= \Zop^{2(k+1)}\,,\quad
{}^{\tau}\! K_{U(1)}^0(SU(2))\,=\, 0\,.
}
Comparison with \fusionringischargelattice\ then yields that
\eqn\Kteisfusion{
K_{U(1)}^1(SU(2)) \,=\, {R_{u(1)}\over I_{2(k+1)}} \,,
}
which is reminiscent of the statement in \refs{\FHTtwo, \FHTone} in the
case of the topological $G/G$ coset and $G$-equivariant twisted
K-theory of $G$. 

\subsec{Comparision between K-theory and D-brane charges}

In the comparison between K-theory and D-brane charges there are two
subtle points that one has to take into account.
Firstly, in order to compare the K-theory to the CFT, we note, that not only the equivariance with respect
to the $u(1)$ action has to be taken into account. Recall that the
selection and identification rules amount to modding out in addition
with the common centre of $su(2)$ and $u(1)$, \ie, with $\Zop_2$ as
specified in the second equation of \cosetrules. Hence, the CFT needs
to be compared to ${}^\tau
K_{U(1)/\Zop_2}(SU(2))$. The key change that occurs in the above K-theory
computation is then that \keytwo\ is replaced by
\eqn\keynotwo{
{}^{\tau}\! K_{U(1)/\Zop_2}^0(S^1)= {}^{\tau}\! K^0(pt.)=  \Zop \,,
}
wherefore 
\eqn\KetsutwoZ{
{}^{\tau}\! K_{U(1)/\Zop_2}^1(SU(2))\,=\, \Zop^{\kappa-1}= \Zop^{k+1}\,,\quad
{}^{\tau}\! K_{U(1)/\Zop_2}^0(SU(2))\,=\, 0\,.
}
This corresponds precisely to the charges carried by the even (or odd)
branes. 

The second point to note is that K-theory is meant to classify the
charges of D-branes in string theory (as opposed to CFT). In addition to conformal
field-theoretical consistency conditions, in string theory the GSO-projection has to be imposed upon the boundary states.
As is
familiar from other super-string backgrounds (see \eg,
\refs{\mrgnonbps}), this does indeed cut down the number of allowed
D-branes. {\it E.g.} in flat space Type II theories, GSO-invariance of
the boundary states implies that both the NSNS and the RR sector
parts of the boundary states have to be GSO-invariant, which implies that there are only half as many GSO-invariant boundary
states than Ishibashi states (as opposed to BCFT, where the number of
Cardy states equals the number of Ishibashi states).
We conjecture that a similar reduction occurs for the above coset
models. In the
present case the NS (R) sector corresponds to even (odd) $a$, \ie, the
NSNS and RR sector branes are precisely the even and odd branes,
respectively. 
To confirm this, one should consider a proper, critical string
background based on these coset theories, such as are provided by Gepner models. 

%%%%

\newsec{$\N=2$ coset models and D-brane charges}

\noindent
In the following, we shall generalize the computations from
$SU(2)/U(1)$ to the Grassmann models and generalized parafermions. 
We will step back for a moment and provide some relevant facts of
these ${\cal N}=2$ coset CFTs. 

\subsec{Supersymmetric coset models}

Recall that any WZW model for a compact Lie group $G$, described by affine ${\g}_k$ at level $k$, can be supersymmetrized to
an $\N=1$ SCFT by introducing dim($\g$) fermionic fields transforming in the adjoint
representation \refs{\KT}. This theory is most elegantly described by
`decoupling' the bosonic and fermionic degrees of freedom (see \eg\
\refs{\KS}), which results in shifting the level of the KM theory. The
$\N=1$ supersymmetric version for $\g_k$ is then
\eqn\susyWZW{
\g_{k-g}\, \oplus\, {so}(dim(\g))_1 \,,
}
where $g$ is the dual Coxeter number such that $f_{acd}f_{bcd}=g\,\delta_{ab}$.
The central charge of the model is thus 
\eqn\csusyWZW{
c\,=\, \half\, dim(\g)\,+ \,{(k-g)\,dim(\g)\over k}\,.
}
Due to the product structure of \susyWZW, it usually suffices to study
the bosonic $\g_k$ model and infer from there the supersymmetric
statements by tensoring with the ${so}(d)_1$ theory. In
particular, this was used in \refs{\FSWZW} to compare the boundary
states to K-theory charges. 

The supersymmetrization of coset models is slightly more intricate. Let $\h$ be a sub-Lie algebra of $\g$. The supersymmetric construction
for $\g_k$ contains one for $\h$, and one can apply the coset
construction to this pair of theories. 
The so-obtained $\N=1$ supersymmetric coset models are of the form
\eqn\susycosets{
{\g_{k}\, \oplus\, {so}(2d)_1 \over \h_{k+g_{\g}-g_{\h}}} \,,
}
where $d=\half (dim(\g)-dim(\h))$ and $g$ denotes the dual Coxeter
numbers of the respective Lie algebras. The ${so}(2d)_1$ part
again parametrizes $d$ complex fermions. Note that this corresponds to
the supersymmetric version of $\g_{k+g_{\g}}/\h$. It is important to
realize that $\h$ is embedded diagonally into $\g_{k}\, \oplus\,
{so}(2d)_1$ and hence, that the theory does not simply factorize into
bosonic and fermionic contributions, as \eg\ in the WZW models. This
is apparent from the way the $\h$ current algebra is constructed in \refs{\KS}. 
One consequence of this for the present purposes is that the bosonic
results for coset branes in $\g_k/\h$ will not simply carry over to
the supersymmetrized theory. There are
non-trivial selection rules for highest weights in the supersymmetric
coset, which non-trivially mix the weights of $\g$, ${so}(2d)$
and $\h$ (see below). 

Generically, \susycosets\ is an $\N=1$ theory. As is well known from
\refs{\KS}, for rank$(\g)=$ rank$(\h)$, the coset theory is $\N=2$ if the geometric
(right-action) coset space $G/H$ of the corresponding Lie groups is
K\"ahler.
The representations of the coset model \susycosets\ are labeled by
highest weights $(\Lambda, \lambda, a)$, which correspond to
$\g_k$, $\h_{k+g_\g-g_\h}$ and ${so}(2d)_1$ highest
weights, respectively. However, not all combinations of weights are distinct in
the coset theory. Gepner developed a general method how to extract the field
identifications in coset models by analyzing the action of certain
automorphisms on the S-modular matrix \refs{\gepnerFI}.

\subsec{Grassmanian coset models $Gr(n,k)$}

The models of interest in
the following are the generalized super-conformal minimal
models based on the Grassmannians $G(n, n+k)$ of $n$-planes in
$\Cop^{n+k}$, which after level-rank duality take the form of
projective cosets
\eqn\smm{
Gr(n,k)\,:\ {{su}(n+1)_{k} \oplus so(2n)_1 \over {su}(n)_{k+1}\oplus {u}(1)_{n(n+1)(k+n+1)}} \,.
}
The level of ${u}(1)_{n(n+1)(k+n+1)}$ indicates the weights are
$\Zop_{n(n+1)(k+n+1)}$-valued. The central charge is $c= 3nk/(k+n+1)$
and the weights are labeled by $(\Lambda,\lambda, m, a)$, where
$\lambda$ and $m$ correspond to $su(n)$ and $u(1)$ weights, 
respectively, as well as $so(2n)_1$ weights $a\in\Zop/4$. 
%%and the ${\cal
%%N}=2$ $U(1)$ charge for a state with highest weight $(\Lambda,
%%\lambda, m, a)$ is given by
%%\eqn\grassUone{
%%Q= \sum_{i=1}^{n} a_i - {m\over k+n+1}\, \hbox{mod}\, 2\Zop \,,
%%}
%%which follows from the representation of the ${\cal N}=2$
%%superconformal algebra in the coset theory as in \refs{\KS}.
The highest weights in the basis of fundamental weights are
\eqn\grassHW{
\Lambda= (\Lambda_1,\cdots, \Lambda_n)\,,\ \lambda=(\Lambda_1, \cdots, \Lambda_{n-1})\,,\
m= (n+1) \Lambda^{(n)}\cdot \Lambda = \sum_{i=1}^n i \Lambda_i \,.
}
Not all combinations of weights in \grassHW\ are
distinct in the theory and certain field identifications need to be
imposed. Recall from \refs{\gepnerFI} that an automorphism $\sigma$ of the extended Dynkin diagram of $\g_k$ is defined to
be {\it proper}, if 
\eqn\properauto{
\sigma(\lambda)\,=\, \sigma(0)\,+\, w_{\sigma}(\lambda)\,,}
for some fixed $w_{\sigma}\in W$.
For any proper automorphism $\sigma(0)=k\lambda_{\sigma}$, where
$\lambda_{\sigma}$ is a level 1 weight (specified for each such $\sigma$). 
The field identification
rules are extracted by considering the action of proper automorphisms on the S-modular
matrix. Let $S^{\g,k}_{\Lambda, \Gamma}$ be the S-modular matrix
for $\g_k$. Using the explicit
form of $S$ it follows that
\eqn\sigmaonS{
S^{\g,k}_{\sigma(\Lambda), \Gamma}\,=\, e^{2\pi i
\lambda_{\sigma}\Gamma}\, S^{\g,k}_{\Lambda, \Gamma}\,.
}
Then, fields for which the phase in \sigmaonS\ vanishes need to be
identified in order to ensure that the modular matrix has full rank. 

Consider thus the proper
automorphisms, acting on ${su}(n+1)$ and ${so}(2d)_1$, respectively, as
\eqn\grassproper{
\sigma(\Lambda)\,=\, k \Lambda^{(1)}\,+\, w_{\sigma}(\Lambda)\,,\quad 
\sigma_v(a)\,=\, v + w_{\sigma_v}(a)\,,
}
where $v$ is the ${so}(2d)_1$ vector weight. Note that for
${su}(n+1)$ the proper outer automorphisms are given by the order $n+1$
rotation group of the extended Dynkin diagram, \ie
\eqn\suauto{
\sigma\left(\sum_{i=1}^n\, \lambda_i\Lambda^{(i)}\right)\,=\, (k-\lambda_1-\cdots-\lambda_n)\,
\Lambda^{(1)}+ \lambda_1 \Lambda^{(2)}\,+\cdots+\lambda_{n-1}\Lambda^{(n)}\,.
}
There are two combinations of these automorphisms that can be applied
to the coset fields
\eqn\fieldids{\eqalign{
J_1\,(\Lambda, \lambda, m, a) \,=\, (\sigma(\Lambda),\, \sigma(\lambda),\,
m+k+n+1,\, \sigma_v(a))\,,\cr
J_2\,(\Lambda, \lambda, m, a) \,=\, (\sigma(\Lambda),\, \lambda,\,
m+n(k+n+1),\, \sigma_v^n(a))\,,
}}
which are of order $n(n+1)$ and $n+1$ respectively, since $\sigma$ on
$su(n)$ is of order $n$.
Applying \sigmaonS\ and \fieldids, shows that the phase vanishes, and thus that the
S-matrix elements for the coset field and its images under $J_1$ and $J_2$
are the same ({\it cf.} \refs{\LVW, \gepnerFI}). This leads us to identify the fields \fieldids\ in order to
maintain unitarity of the S-matrix. 
For later reference, recall that the dimension of the chiral ring is
$$
\hbox{dim}\, {\cal R}^{Gr(n,k)} = {(n+k)!\over n!\, k!}\,.
$$

A peculiar property of the $Gr(n, k)$ is that they are conjectured
to be invariant under the exchange of $k$ and $n$, \ie, 
\eqn\levelrank{
Gr(n,k) \cong Gr(k, n) \,.
}
Despite the lack of a full proof of this level-rank duality, very
convincing arguments have been put forward, most notably in \refs{\KS},
where it is shown that the coset representation
of the ${\cal N}=2$ superconformal algebra is symmetric in $k$ and $n$. Furthermore, in \refs{\schnitzerlevelrank} an isomorphism between
the chiral data is constructed. This implies in
particular, that the corresponding boundary theories need to respect level-rank duality.

The boundary intersection index $\I_{ab} = \Tr_{\H_{ab}^{open}} (-1)^F
= {}_{RR}\langle\!\langle a|\!|b\rangle\!\rangle_{RR}$
\refs{\douglasfiol, \BDLR, \horivafa} for the Grassmannian coset models has been
computed in \refs{\LW}. It was indicated there that, as discussed earlier for the
superminimal model, it is sufficient to compute the index for the $\Lambda=0$ Cardy boundary states.
%% and is best described by the structure constants of 
%%\eqn\grbfr{
%%{\cal O} = {R_{su(n)\oplus u(1)}\over I_{k, n(k+n)}}\,.
%%}
%%In the case of $n=1$
%%this correctly reproduces that $I= \langle \zeta^{(k+1)} \rangle$
%%(for, say, the even branes). 
The rank of the charge lattice is equal to the dimension of the chiral
ring, \ie, 
\eqn\rankI{
\left(\matrix{n+k \cr n}\right)\,.
}
The result \rankI\ refers to the
charges of even or odd maximally supersymmetric D-branes in the
Grassmannian coset CFTs and the purpose of the subsequent section is to
derive this result for the charge lattice by a K-theory computation.

%%%%

\subsec{Superparafermion coset modesl $SPF(n,k)$}

A related coset model is that of $SU(n)$ by its maximal torus, \ie\ the
generalized superparafermions, 
\eqn\smt{
SPF(n,k)\,:\  {{su}(n+1)_{k}\oplus so(n(n+1))_1 \over u(1)^{n}}\,.
}
The BCFT will be discussed elsewhere, however, we shall
provide the corresponding K-theory calculation below. The selection
and identification rules which are needed for this computation are as follows. 

The center of
$SU(n+1)$,
or equivalently the outer automorphism group of its affine version, is generated by the order $n+1$ rotation $\sigma$ of the extended Dynkin
diagram. Again, $\sigma_v$ acts on the $so(n(n+1))_1$, with
$\sigma_v(0)=v$. This imposes the identification rule
\eqn\parafieldid{
\left(\,\sigma(\Lambda)\,,\, {\bf m}+\sigma(0)\,,\, \sigma_v^n(a)\,\right)\,=\, 
(\Lambda\,,\,{\bf m}\,,\,a\,)\,,
}
where $\sigma(0)= (k+n+1) \Lambda^{(1)}$.

The dimension of the chiral ring
in this case, using \refs{\LVW}, is computed to be
$$
\hbox{dim}\, {\cal R}^{SPF(n,k)} = {(k+n)!\over k!}\,.
$$

%%%%%%

\newsec{Twisted equivariant K-theory}

\noindent
This section provides the computation of the twisted equivariant
K-groups relevant for the generalized parafermions and Grassmannian coset
models.

\subsec{${}^{\tau}\! K_{H}(G)$}

Let $G$ be a simple, simply-connected, connected Lie
group (\ie, in particular $\pi_1$ is torsion-free). For most purposes we shall
consider $G=SU(N+1)$. Further let $H$ be a connected, maximal rank subgroup of
$G$. Then the computation of ${}^{\tau}\! K_{H}(G)$ can be traced back to the
computation of Freed, Hopkins and Teleman in \refs{\FHTtwo,\FHTthree}
of ${}^{\tau}\! K_{G}(G)$. The theorem by FHT states, that twisted $G$-equivariant
K-theory of $G$ is isomophic as an algebra to the Verlinde algebra of
$G$ at a level $k$ determined by the 3-form that specifies the twist
$\tau$ of
the K-theory
\eqn\FHTtheorem{
{}^{\tau}\! K_{G}^{dim(G)}(G)\,=\, V_k(G)\,.
}
The twist $\tau$ given by the $k+g$ multiple of the generator of
$H^3(G,\Zop)$, which corresponds to the ${\bf H}=dB$ 3-form flux.
The important point about \FHTtheorem\ that stresses its possible
relevance for coset models is that the $G$-equivariance is taken with
respect to the adjoint action of $G$ on itself.

The case of coset models, where $H$ is a proper, connected, maximal rank subgroup,
relates to this result as follows. First define
\eqn\ghg{
G^c\times_H G = \left\{ (g, \tilde{g})\in G\times G\,; \ 
(g, \tilde{g}) \sim (h^{-1} g h, \tilde{g}h)\,, h\in H    \right\}\,.
}
Here, $G^c$ indicates that the action of $H$ on $G$ is taken to be the
conjugation action, whereas on the second factor, the action is the
right action. Then the map
\eqn\phiiso{\eqalign{
\Phi\, : \quad G^c \times_H G \ &\rightarrow\  G^c \times G/H\cr
(g, \tilde{g}) \ &\mapsto (\tilde{g} g \tilde{g}^{-1}, \tilde{g} H)\,,
}}
is well-defined and an isomorphism. In order to calculate the twisted
equivariant K-groups, note further, that for any $H$-space 
\eqn\Hspacerel{
K_G(X\times_H G)\cong K_H(X)\,,
}
where the equivariance is with respect to the specified action of $H$
on $X$. This is simply the statement that any $G$-bundle over $G\times_{H} X$
is determined by the underlying $H$-bundle over $X$. Hence, setting $G^c=X$ we obtain
\eqn\KGHcalculation{
K_G(G^c\, \times\,  G/H)= K_G(G^c\, \times_H \, G)  \cong K_H(G^c)\,,
}
For $\pi_1(G)$ torsion-free, an equivariant form of the K\"unneth
formula holds by a result of Hodgkin and Snaith
\refs{\hodgkin, \snaith}, whereby\footnote{$^{\flat}$}{More precisely, the theorem
states the existence of a K\"unneth spectral sequence, with
$E_2^{-p,q}= Tor_{R_G}^{p}({}^{\tau}\! K^q_{G}(G) , K^q_H(pt.))$ converging to
${}^\tau K_H(G)$.
However, in the present context $R_H$ is a free module over $R_G$,
wherefore the higher Tor's vanish
and the spectral sequence collapses. See \eg\
\refs{\volkerK} for a case where this spectral sequence is
non-trivial.}
\eqn\KGHcontinued{
{}^{\tau}\! K_{H}(G)\,=\,{}^{\tau}\! K_{G}(G)\, \otimes_{R_G}R_H\,,
} 
where $R$ denotes as before the representation rings of the respective
groups. Note that we used the isomorphism $K_{G}(G/H) \cong
K_{H}(pt.)$, which is true since any $G$-bundle on $G/H$ is of the
form $G \times_{H} R \,\arrow\, G \times_{H} pt. = G/H$ for a
$H$-representation $R$.

We can now complete our computation of $K_H(G)$. For $R_H$ free as an $R_G$-module, which holds for $H$
of maximal rank and connected \refs{\pittie}, we can use the FHT
result \FHTtheorem\ to simplify \KGHcontinued\ into
\eqn\KGHfinal{
{}^{\tau}\! K_{H}(G)\,=\, {\, R_H\, \over I_G}\,.
}
$I_G$ is the Verlinde ideal of $G$, that is, the
ideal, that provides the identifications necessary in order to obtain
the Verlinde algebra at a specified level from the representation ring of $G$
as $V^G = R_G/I_G$.  
The Schubert-like combinatorial relations, that define the ideal
$I_G$ were obtained in \refs{\gepnerFRG}. 

First, in order to elucidate \KGHfinal, it is useful to recall how the representation
rings of $G$ and $H$ are related. Since any $G$-representation can be
decomposed into $H$-representations, $R_G< R_H$. More precisely,
by definition the representation ring is
\eqn\repring{
R_G\,\cong \, \Zop [M^\star ]^{W_G} \cong \Zop [\Lambda^{(1)},\cdots, \Lambda^{(n)}]\,,
}
where $M^\star$ is the weight lattice, $W_G$ the Weyl group, so that $
\Zop [M^\star ]^{W_G}$ is the polynomial ring of invariant polynomials
in the weights. Since any representation can be decomposed into
representations with highest weights equal to the fundamental weights
$\Lambda^{(i)}$, the latter form a basis of the representation
ring. 

%%%%

\subsec{Twist-class and level of the Verlinde-ideal}

To compare the above results with the intersection index calculation
it is interesting to
compute the induced level of the Verlinde ideal entering the K-theory
calculation in \KGHfinal. 
Recall, that each invariant symmetric bilinear form on the Lie algebra $\g$ is in
one to one correspondence with central extensions 
$\Rop\,\arrow\,\widetilde{L\g}\,\arrow L\g$ of the loop algebra
\refs{\PS}. This relation follows by noting that each such
bilinear form $\langle ,\rangle$ on $\g$ defines a cocycle $\omega:\, L\g \times L\g
\arrow \Rop$ on the corresponding loop
algebra $L\g$ as
\eqn\omegacocycle{
\omega (\xi(z), \eta(z))\,=\, \hbox{Res}_{z=0}  \ \langle \,\xi(z),\hbox{d} \eta(z)\,\rangle\,.
}
If $\omega$ is integral then there is a corresponding central
extension $U(1)\arrow \widetilde{LG}\arrow LG$ of the loop group $LG$. 

In the case of ${}^\tau K_G(G)$, the twist-class is an equivariant integral
cohomology class $\tau$ in $H^3_G(G)$. In \refs{\FHTone} the twistings were
chosen, such that the restriction to $H^3_T(T)$, where $T$ is the
maximal torus of $G$, have trivial components in $H^3(T)$. In this
case it was shown that the twisted equivariant K-theory with complex
coefficients can be computed as a twisted cohomology with coefficients in a
line-bundle ${\cal L}^\tau$, which is determined by the class $\tau$,
\eqn\fhtcomplex{
{}^\tau K_G^{0/1}(G; \Cop)^\wedge_g \,=\, {}^\tau H^{even/odd} (G^g;
{\cal L}^\tau (g)) \,,
}
where ${}^\wedge$ denotes completion (see \eg\ \refs{\Hartshorn}). 
${\cal L}^{\tau}$ is determined by $\tau$ as follows. $\tau \in H^2
(BT) \otimes H^1 (T)$, by the assumption that
$\tau|_{H^3(T)}=0$. Thus, it defines a map $\tau : T\rightarrow
T^\vee$. The sheaf ${\cal L}^\tau (t)$ for $t \in T$ is $\tau(t)$,
which specifies a line bundle over $T$, since $T^\vee$ labels
line-bundles over $T$.
${}^\tau K_G^{0/1}(G; \Cop)^\wedge_g$ can be viewed as the completion of sections of a
sheaf (denoted ${}^\tau {\cal K}$ in \refs{\FHTone}) over the
GIT-quotient $G_{\Cop}/\!/G_{\Cop}$. The simplification that occurs in the above
$K_G(G)$ case is that ${}^\tau {\cal K}$ is a skyscraper-sheaf
with one-dimensional stalks, and thus the twisted K-theory can 
be calculated exactly by means of this completion technique. Note, that
these results hold for K-theory with complex coefficients, and thus in
particular do not take into account torsion. The case of integral
K-theory will be discussed in \refs{\FHTintegral}.

All this extends to the case of interest in this paper, \ie, $H$-equivariant K-theory of $G$.
The twistings are in $H^3_H(G)$, and we restrict them to
$H^3_{T_H}(T_G)$, whereby $T_G$ is the maximal torus of $G$. Again, in 
$H^3_{T_H}(T_G)= H^3(ET_H \times_{T_H} T_G)$ we assume that $\tau$
restricts trivially to $H^3(T_H)$. Thus $\tau$ gives rise to a map
$T_H\arrow T_H^\vee$, thus specifying a line-bundle over each point in
$T_H$.

A comment on induced level. Fix an element $\tau\in H^3_H(G)$ to be the (topological) level of
a central extension of $LG$. Then we can write this as $k+g$ times the
generator of $H^3_H(G)$, with $k$ the level and $g$ the dual Coxeter
number. The induced twisting on a subgroup $H$ will be $k+g$, wherefore the induced level
is $k+g-g_H$, where $g_H$ is the dual Coxeter number for $H$. 

%%%%

\subsec{$K_{U(1)}(SU(2))$ revisited }

As a first application of the above results, we revisit the calculation for $SU(2)/U(1)$, which we
already performed using the cell complex earlier. The representation
ring of $SU(2)$ is $R_{SU(2)}= \Zop [\Lambda]$, where $\Lambda$ labels the
fundamental representation. The ring $R_{U(1)}$ is generated
by the one-dimensional representation. More precisely, a
representation of $U(1)$ is specified by an integer $p$, such that
\eqn\Uonerep{\eqalign{
\zeta^p:\ U(1)  &\arrow GL(\Cop)\cr
e^{i\theta} & \mapsto e^{ip \theta} \cdot\qquad ,
}}
\ie, $U(1)$ acts by multiplication with the $p$th powers of its
generators. Since $\zeta^p\otimes \zeta^q = \zeta^{p+q}$, the
representation ring is generated by $\zeta, \zeta^{-1}$ and is isomorphic to
$\Zop [\zeta, \zeta^{-1} ]$.
The twisting for $SU(2)$ is $k+2$ and the standard Verlinde ideal
is\footnote{$^\sharp$}{In the following, only the $\Zop$-module
structure is taken into account, since our main interest is the rank
of the charge lattice, rather than the ring structure.} ${I}^k_{SU(2)} = \langle \Lambda^{k+1}
\rangle$. To compute \KGHfinal\ in this context we note that the
representation ring of $SU(2)$ is embedded into that of $U(1)$ by the
identification $\Lambda=\zeta+\zeta^{-1}$, and thus
\eqn\Ksutwofinal{
{}^{\tau}K_{U(1)}(SU(2)) 
= {R_{SU(2)}\over \langle \Lambda^{k+1} \rangle}\otimes  R_{U(1)} 
= {\Zop(\zeta) \over \langle \, (\zeta+ \zeta^{-1})^{k+1}\,  \rangle } = \Zop^{2(k+1)}\,,
}
in degree $dim (SU(2))$, \ie\ for all odd degrees, and vanishes
otherwise. This is precisely our result \Ketsutwo. The induced level for the Verlinde ideal as embedded into
$R_{U(1)}$ is again $k$ and the twisting $\tau= (k+2)$.

%%%%

\subsec{$K_{U(n)}(SU(n+1))$}

Next we generalize this to the Grassmannian coset models. First recall
that 
\eqn\dimchiral{
V_{SU(n+1)}^{k} \cong \Zop^{n+k\choose k} \,.
}
The twisting
in $SU(n)$ is $k+n+1$ times the generator of $H^3_{H}(G)$. By the
reasoning in the last section, the twisting for the $SU(n)$ subgroup
is thus also $k+n+1$, which corresponds to the level $k+1$ in
$SU(n)$. \KGHfinal\ implies that 
\eqn\grassK{
{}^\tau K_{U(n)}(SU(n+1)) = { R_{U(n)} \over I_{SU(n+1)}^k}
\,,}
in degree $dim (SU(n+1))$ and vanishes for the complementary degrees. 
This can be made more explicit as follows. From Theorem 3 in
\refs{\pittie} we know that for a maximal rank subgroup, which is
connected and closed
\eqn\RH{
K(G/H) \cong R_H \otimes_{R_G} \Zop \,,
}
where $G/H$ is the right quotient. Applied to the present situation,
where $SU(n+1)/U(n)= \Cop P^n$, and using the K-theory for
complex projective space as computed by Adams in \refs{\adams}, this reads
\eqn\KCPn{
R_H \otimes_{R_G} \Zop \cong K(\Cop P^n) \cong \Zop^{n+1}\,.
}
Together with \dimchiral, \grassK\ simplifies therefore to
\eqn\GrassKfinal{
{}^\tau K_{U(n)}(SU(n+1)) \cong \Zop^{(n+1) \,{n+k\choose k}}\,.
}

\subsec{$K_{U(1)^n}(SU(n+1))$}

For generalized super-parafermion cosets, we invoke Theorem 1 in
\refs{\pittie}, which implies, that $R_{u(1)^n}$ is free of rank $(n+1)!$, \ie, 
\eqn\Rmaxtorus{
R_{U(1)^n} \cong \left( R_{SU(n+1)} \right)^{|W|} = \left( R_{SU(n+1)} \right)^{(n+1)!}\,,  
}
where $|W|$ is the order of the Weyl group. Therefore \KGHfinal\ results in
\eqn\KGPfinal{
{}^\tau K_{U(1)^n} (SU(n+1)) =  \left({R_{SU(n+1)}\over
I_{SU(n+1)}^k}\right)^{\oplus (n+1)!} = \Zop^{(n+1)! {n+k\choose k}}\,.
}

\subsec{Comparison between K-theory and D-brane charges}

To compare the above K-theoretical results to the charge lattice that
the boundary conformal field theory and string theory predicts, there are again two
subtle points that need to be elucidated. Firstly, as in the $SU(2)/U(1)$ case, the identification rules implied
by the order $n(n+1)$ outer automorphism \suauto\ necessitate that the
K-theory has to be computed with respect to an additional equivariance.
The automorphisms \fieldids\ act on the affine $su(n+1)$ weights as
outer automorphisms, and thus correspond to central elements of the
corresponding Lie group $SU(n+1)$. 
For this reason, the equivariance in the K-theory for $Gr(n,k)$ is with respect to $U(n)/Z$, where
\eqn\Zcenter{
Z = \Zop_{ n (n+1) } = \left\langle 
e^{2\pi i/(n+1)} \ \one_{n+1} \,, \quad
\pmatrix{ e^{2\pi i /n } \ \one_n & 0 \cr 0 & 1} 
\right\rangle\,,
}
where the first element generates the center of $SU(n+1)$ and the
second generates the center of $SU(n)$. This implies
\eqn\HmodZ{
H/Z = U(n)/ \Zop_{n(n+1)} = \left.{SU(n) \times U(1) \over \Zop_{n}}
\right/ \Zop_{n(n+1)} = U(n) / \Zop_{n+1} \,.
}
Therefore there is an additional $\Zop_{n+1}$ identification that
needs to be taken into account in the K-theory computation, in order
to compare with the boundary conformal field theory. Now note that $\Zop_{n+1}$ sits inside $U(1)
\subset U(n)$, so that the condition on the subgroup being connected
still holds. That the above arguments goes
through for the $H/Z$-equivariant K-theory\footnote{$^{\natural}$}{I thank S. Fredenhagen for pointing out some subtleties
in this argument.}, is seen most directly in the
formulation of \refs{\FHTone}. Recall from section 4.2, that the twisted
equivariant K-theory ${}^\tau K_H(G)$ can be computed by taking
sections of a sheaf ${}^\tau {\cal K}(G)$,
which is defined on $H_{\Cop}/\!/H_{\Cop}$. There is a natural
$Z$-action on $H_{\Cop}/\!/H_{\Cop}$, and in order to compute the
$H/Z$-equivariant K-theory, one again takes sections of ${}^\tau {\cal
K}(G)$, whose stalks are still 1-dimensional, however, the
$Z$-action forces to take invariant sections, which amounts to
identifying the stalks over the points in each orbit of $Z$. 
This leads for the present case to an $(n+1)$-fold identification
of the stalks, and thus, the K-groups, \ie, the sections of the sheaf
in question, are computed to be 
\eqn\KHZG{
{}^\tau K_{U(n)/\Zop_{n+1}}(SU(n+1)) = \Zop^{n+k\choose k} \,.
}  
Note that this is in complete agreement with the charges obtained from
the even (odd) Cardy branes in the CFT. Further, it is consistent with level-rank
duality. 

Analogously, for the generalized superparafermion cosets, the
automorphism is of order $n+1$ and similar reasoning implies that the
charge lattice is again given by the dimension of the chiral ring
\eqn\KHZGpara{
{}^\tau K_{U(1)^n/\Zop_{n+1}}(SU(n+1)) = \Zop^{n !\, {n+k\choose k}} \,.
}
It would be interesting to check this on the CFT side. 

The second point that needs to be addressed refers to the D-brane charges as one would expect
them from string theory. As discussed in the case of $SU(2)/U(1)$,
a full string theoretical treatment would require to GSO project,
which in particular necessitates to cut down the D-brane spectrum to
particular superpositions of RR and NSNS sector boundary states. This
may serve as an explanation, why the K-theory sees only half of the
brane charges of the CFT.

%%%%

\newsec{Conclusions and outlook}

\noindent
In this paper we computed twisted equivariant K-groups, as they arise
in the context of Kazama-Suzuki models. In the case of Grassmannian
coset models the charges obtained from the K-theory match with
the conformal field theory charges, as obtained from even (or odd) A-type
Cardy states. The
K-theory is obtained as a quotient of the representation
ring of the denominator group. In
particular, the charge lattice satisfies the level-rank
duality of the bulk theory. Unlike the WZW models,
where the twisted K-theory for $SU(n+1)$, $n\geq 3$, is strictly larger
than the charge lattice of Cardy branes, in the coset model case, the
Cardy brane charge lattice (at least for the Grassmannian cosets) is
precisely given by the twisted equivariant K-theory. However, as we
pointed out, there is a subtlety regarding the comparison to D-brane
charges in string theory, which seems to be rooted in the
GSO-projection. 

It would be an interesting check to compute the
CFT intersection index in the case of the generalized
parafermions, but we shall postpone this to elsewhere. In view of the
issue related to the GSO-projection, it would be interesting to extend the above
K-theoretical computations to \eg, Gepner models, where an unambiguous
string theory description is at hand, and to compare them to the
charges computed in \refs{\horivafa, \BDLR}. 
More generally, the FHT theorem may be of help in order to shed some more light on the
relation between BCFT data and twisted K-theory.

%%%%

\vskip 1cm

\centerline{{\bf Acknowledgments}}
\pano
It is a great pleasure to thank Constantin Teleman, for many
illuminating discussions on K-theory, as well as Matthias
Gaberdiel and Greg Moore. I am also very grateful to Volker Braun, Stefan
Fredenhagen, Peter Goddard, Andreas Recknagel, and Daniel Roggenkamp for
discussions at various stages of this project. I thank St John's College,
Cambridge, for a Jenkins Scholarship and acknowledge partial support by EEC contract HPRN-2000-00122.

%%%%

\listrefs

\bye